\documentclass[twocolumn,showpacs,preprintnumbers,amsmath,amssymb]{revtex4}

\usepackage{graphicx}
\usepackage{hhline}
\usepackage{dcolumn}
\usepackage{bm}

\usepackage[dvips]{color}

\newcommand{\bluecom}[1]{}

\begin{document}

\title{Critical Phase of Bond Percolations on Growing Networks}

\author{Takehisa Hasegawa}
\email{hasegawa@stat.t.u-tokyo.ac.jp} 
\affiliation{%
Graduate School of Information Science and Technology, The University of Tokyo,
7-3-1, Hongo, Bunkyo-ku, Tokyo, JAPAN.
}%
\author{Koji Nemoto}
\email{nemoto@statphys.sci.hokudai.ac.jp}
\affiliation{%
Department of Physics, Graduate School of Science,
Hokkaido University, Kita 10-jo Nisi 8-tyome, Sapporo, JAPAN.
}%


\begin{abstract}
The critical phase of bond percolation on the random growing tree is examined.
It is shown that the root cluster grows with the system size $N$ as $N^\psi$ and
the mean number of clusters with size $s$ per node 
follows a power function $n_s \propto s^{-\tau}$ 
in the whole range of open bond probability $p$.
The exponent $\tau$ and the fractal exponent $\psi$ are also derived
as a function of $p$ and the degree exponent $\gamma$, and are found to
satisfy the scaling relation $\tau=1+\psi^{-1}$.
Numerical results with several network sizes are quite well fitted 
by a finite size scaling for a wide range of $p$ and $\gamma$,
which gives a clear evidence for the existence of a critical phase.
\end{abstract}
\maketitle

\section{introduction}
The study of complex networks has been one of the most popular topics for many research fields in the last decade 
\cite{BarabasiRev, NewmanRev, BoccalettiRev}.
This activity has been drown by
the discoveries of the {\it small-world} \cite{SWnet} and the {\it scale free} (SF) \cite{SFnet} properties,
 which are common to many real networks, 
e.g., Internet, WWW, social networks, food-webs.
The former means 
the (sub-)logarithmic dependence of the mean shortest distance 
$\bar{l}$ between nodes with network size 
$N$ as $\bar{l} \propto \log N$ ($\propto \log N/\log \log N$) \cite{NewmanRev,SWnet,SWref1}, 
and the latter a power-law tail in the degree distribution 
$P(k) \propto k^{-\gamma}$, 
where the degree $k$ is the number of edges connected to a node.
The dynamics on complex networks, 
such as percolation, epidemic processes, interacting spin systems, coupled oscillators, 
have been extensively studied
with stimulating our interests for
the relationships between network topology and critical phenomena \cite{DorogoRev}. 
In most analytical approaches {\it locally tree-like approximation} is used to 
give many detailed physical pictures about critical phenomena on some network models, 
particularly, on uncorrelated SF networks \cite{DorogoRev}. 
On the other hand, 
many works have reported that systems on some growing networks 
show quite different phase transitions \cite{Callaway,Dorogo01,Kiss03, Coulomb,Bauer,Berker06,Kajeh}.
As shown in \cite{Coulomb,Dorogo01}, 
the percolation on some growing network models undergoes 
an {\it infinite order transition} with a {\it BKT-like singularity}; 
(i) the relative size of the giant component vanishes in an essentially singular way at the transition point, 
so that the transition is of infinite order,
and (ii) 
the mean number $n_s$ of clusters with size $s$ per node (or the cluster size distribution in short)
decays in a power-law fashion with $s$
in the whole region where no giant component exists.
Similar transitions are observed for the interacting spin systems 
on a hierarchical SF network \cite{Berker06}, and 
an inhomogeneous growing network \cite{Bauer,Kajeh}.

The above unusual disordered phase is thought to be the same as 
the critical phase \cite{Berker09}, which is also called {\it patchy} phase \cite{Boettcher}, 
observed on hierarchical SF networks and the Hanoi networks.
By renormalization group (RG) techniques, Berker et al. \cite{Berker09} have studied bond percolations
on the decorated (2,2)-flower \cite{Rozen}, which is one of hierarchical SF networks, to show the existence of a critical phase, 
where RG flow converges onto the line of nontrivial stable fixed points.

Let us turn our eyes upon exotic but regular graphs for a while. 
Critical phenomena on {\it nonamenable graphs} (NAGs), 
which are defined as graphs with positive Cheeger constant, have been studied in recent years \cite{Lyons00, Schonmann01}. 
Roughly speaking, NAGs are regular graphs having small-world property $\bar{l} \propto \log N$.
Hyperbolic lattices and regular trees are typical examples of NAGs. 
It has been predicted that 
the bond percolation on a NAG exhibits a {\it multiple phase transition} (MPT) 
which takes three distinct phases
according to the open bond probability $p$ as follows; 
(i) the non-percolating phase ($0 \le p < p_{c1}$) in which only finite size clusters exist, 
(ii) the critical phase ($p_{c1} \le p \le p_{c2}$) in which there are infinitely many infinite clusters, 
and (iii) the percolating phase ($p_{c2} < p \le 1$) in which the system has a unique infinite cluster.
Here {\it infinite cluster} means a cluster whose mass diverges with system size $N$ 
as $N^\phi$ with $0<\phi \le 1$. 
Recent paper \cite{NAG,NAG2} (see also \cite{Baek, Baek2}) performed Monte-Carlo simulations for 
the bond percolation on the enhanced binary tree, which is one of NAGs, 
to show that the system undergoes the MPT, 
and the critical phase has the following properties; 
(i) continuously increasing of the fractal exponent $\psi$ with $p$, 
where $\psi$ is defined as 
\begin{equation}
s_0(N) \propto N^\psi, \label{s0-psi}
\end{equation}
$s_0(N)$ being the mean size of cluster to which the root node belongs (hereafter we refer to this cluster as the root cluster),
and 
(ii) the cluster size distribution $n_s$ 
always having a power-law tail
\begin{equation}
n_s \propto s^{-\tau}, \label{nspower}
\end{equation}
with $p$-dependent $\tau$ satisfying
\begin{equation}
\tau =1+\psi^{-1} \label{relation}.
\end{equation}
We already know that 
standard systems on the Euclidean lattices 
(which are amenable) have just one critical {\it point} 
and no critical {\it phase} \cite{BK89}, 
while it is unclear and rarely discussed so far
whether percolations or other processes on complex networks have such a critical {\it phase} or not.

The aim of this paper is to connect two concepts,
unusual phase transitions on growing networks and the MPT on NAGs.
For this purpose we analyze the bond percolation on the {\it growing random tree} (GR tree) \cite{KR00,Det00}.
As already mentioned, growing networks are expected to show 
a criticality in the region where its order parameter takes zero, 
while it is also trivial that any tree has no ordered phase due to the absence of loops.
Thus, 
systems on the GR tree are always critical in the whole range of $p$ (except $p=0,1$).
We show analytically that 
the exponents $\tau$ and $\psi$ change continuously with the open bond probability $p$, 
with satisfying the scaling relation (\ref{relation}).
We also perform the Monte-Carlo simulation to show that finite size scaling for $n_s$ is quite well fitted 
for any $p$, irrespective of the degree exponent $\gamma$.

\section{model}
\begin{figure} 
\begin{center}
\includegraphics[width=4cm]{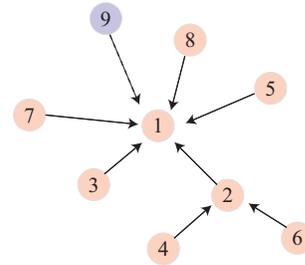}
\end{center}
\caption{
Example of growth of the growing random tree.}
\label{NetGrowth}
\end{figure}

A realization T$_N$ of the GR tree with $N$ nodes is
obtained as a tree at time $t=N$ generated by a stochastic process as follows:
One starts at time $t=1$ with T$_1$ consisting of just one isolated node which we call the initial node or the root (for later convenience it is connected to a dangling bond so that the initial degree is $k=1$). 
At each time step, 
a new node is added and linked to T$_N$ to make T$_{N+1}$ (see figure \ref{NetGrowth}).
The probability that the new node 
is linked to a node of T$_N$ with the degree $k$ is proportional to 
the linear attachment kernel $A_k=k+\alpha$ ($\alpha>-1$).
The stationary degree distribution $P(k)$ of the resulting tree is known to have 
a power-law tail $P(k) \propto k^{-\gamma}$, 
where the degree exponent $\gamma$ is related to $\alpha$ as $\gamma =3+\alpha$ \cite{KR00}.
%

We now consider the bond percolation on the GR tree with $N$ nodes. 
Each bond is open with probability $p$, otherwise closed. 
\section{derivation of $\psi$ and $\tau$}
First, we evaluate the fractal exponent $\psi$ of 
the root cluster. 
Our calculation here is completely in the same way as for the Ising spin system \cite{HN4}. 
Let us assign a level to each node
according to the distance $l$ from the initial node. 
The initial node is regarded as at the $0$-th level.
Let $n_N^{(l)}$ denote the mean number of nodes at the $l$-th level on T$_N$.
Then the sum of all degrees of the nodes at the $l$-th level is equal to $n_N^{(l)}+ n_N^{(l+1)}$.
In adding a new node to T$_N$,
the probability that the new node is linked to any node at the $l$-th level is
$(n_N^{(l)}+n_N^{(l+1)}+\alpha n_N^{(l)})/[(2+\alpha)N-1]$.
If this happens the new node itself is then at the $(l+1)$-th level, 
so we obtain 
\begin{equation}
n_{N+1}^{(l+1)}=n_N^{(l+1)}+\frac{ c_1 n_N^{(l)}+n_N^{(l+1)}}{ c_2 N-1} \quad (l \ge 0),  \label{nNlrec}
\end{equation}
where 
$c_1=1+\alpha$, $c_2=2+\alpha$, 
and the boundary condition is $n_N^{(0)}=1$ for all $N$.
The generating function of $n_N^{(l)}$,
\begin{equation}
G_N(p)=\sum_{l=0}^\infty n_N^{(l)}p^l,
\end{equation}
is nothing but the mean size of the root cluster, 
and the recursion relation is obtained from (\ref{nNlrec}) as
\begin{equation}
(c_2 N-1)G_{N+1}(p)=(c_2 N+c_1 p)G_N(p) -1.
\end{equation}
It is easily solved as 
\begin{eqnarray}
G_N(p)  
&=& \frac{1}{1+c_1p}\nonumber\\
&&+
\frac{c_1p}{1+c_1p} 
\frac{\Gamma(1-{c_2}^{-1})\Gamma(N+{c_2}^{-1} c_1 p)}{\Gamma(1+{c_2}^{-1} c_1 p)\Gamma(N-{c_2}^{-1})}.
\end{eqnarray}%
For $N \gg 1$, the second term becomes dominant,
so that the mean size of the root cluster grows asymptotically as
\begin{eqnarray}
G_N(p) \simeq N^{\frac{c_1 p +1}{c_2}} =N^{\frac{1+(1+\alpha)p}{2+\alpha}},  \label{initialnode}
\end{eqnarray}
and thus we find the fractal exponent $\psi$ as
\begin{equation}
\psi= \frac{1+(1+\alpha)p}{2+\alpha} =\frac{1+(\gamma -2)p}{\gamma-1}. \label{psi}
\end{equation}

\begin{figure}
\begin{center}
\includegraphics[width=7cm]{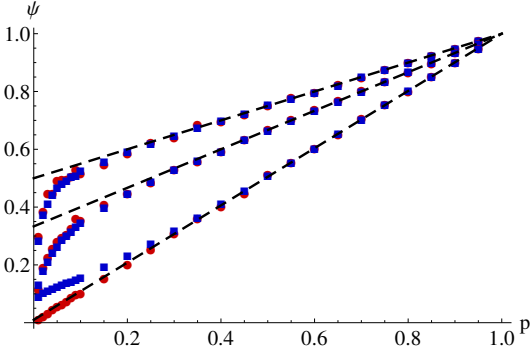}
\end{center}
\caption{
(Color online) Fractal exponent $\psi_{\rm L}$ of the largest clusters (blue square)
and $\psi$ of the root cluster (red circle) 
on the GR tree with $\alpha=0, 1, 100$, from top to bottom.
The dotted lines show the analytical prediction (\ref{psi}).
$\psi_{\rm L}$ and $\psi$ are given by the fit of those with several sizes.
}
\label{psiplot}
\end{figure}


\begin{figure}[t]
\begin{center}
\includegraphics[width=8cm]{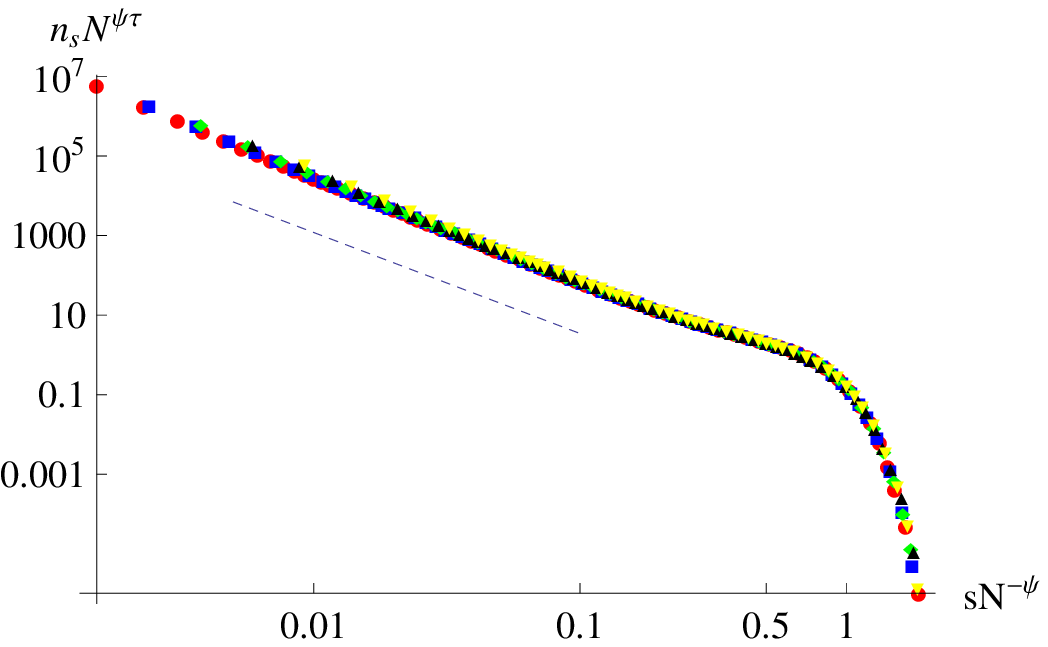}
\includegraphics[width=8cm]{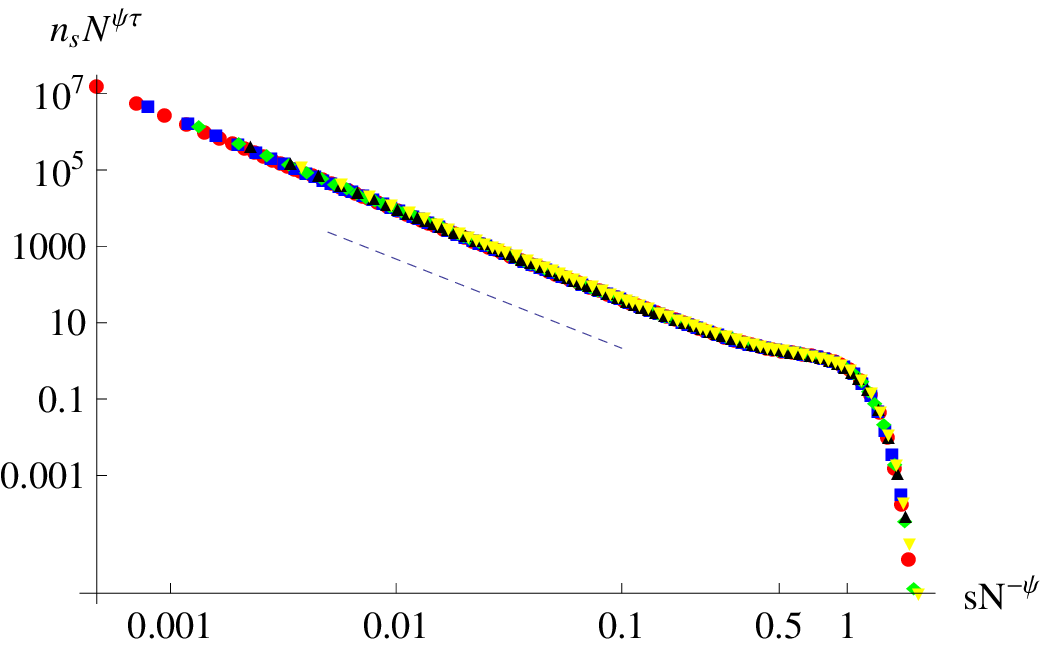}
\includegraphics[width=8cm]{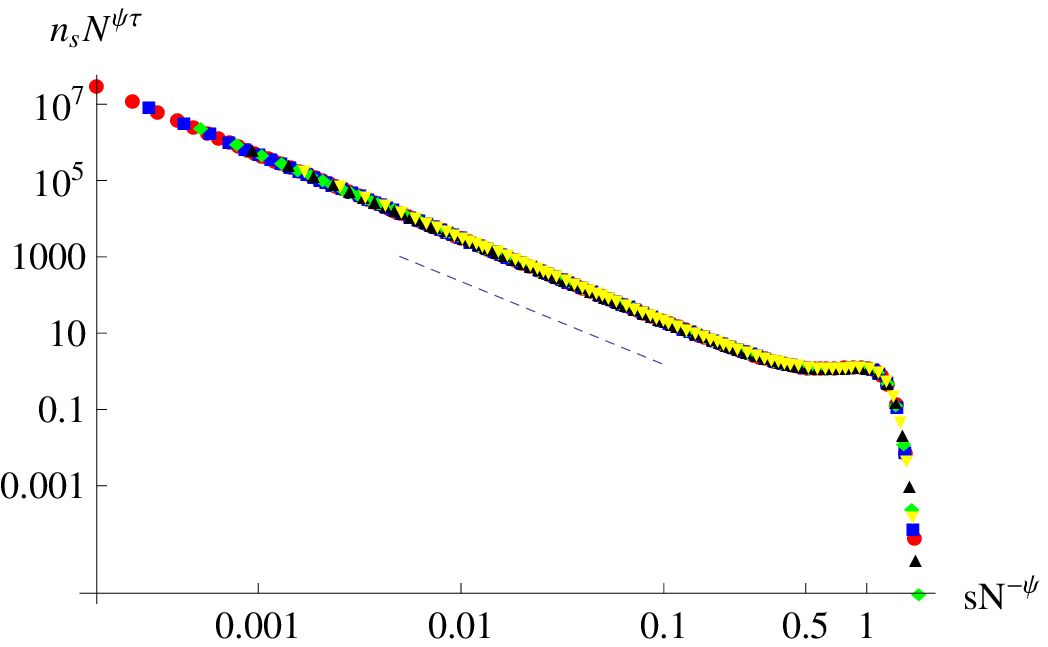}
\end{center}
\caption{
(Color online) Finite size scaling for the cluster size distribution $n_s$ 
of the GR tree with $\alpha =0$, and $p=0.3, 0.5, 0.7$, from top to bottom.
The number of nodes is taken 
$65536$(red circle), $32768$(blue square), $16384$(green diamond), $8192$(black triangle), $4096$(yellow inverted triangle).
The dotted line follows the exponent of Eq.(\ref{tau}).
We obtain $\psi$ by numerical simulations, 
and determine $\tau$ by scaling relation (\ref{relation}) with numerically obtained $\psi$.
}
\label{FSSa0}
\end{figure}


Next, we derive the cluster size distribution $n_s$.
We consider the time evolution of the number $N_s^l(t)$ of clusters 
which have $s-1$ open bonds ($s$ connecting nodes) and $l$ closed bonds at time $t$.
The time evolution of $N_s^l(t)$ is given as
\begin{eqnarray}
N_s^l (t+1) &=& N_s^l(t)
-\frac{\alpha' s +l-2}{\alpha' N-1}N_s^l (t)\nonumber\\
&&+p \frac{\alpha' (s-1)+l-2}{\alpha' N-1}N_{s-1}^l (t) \nonumber\\
&&+q \frac{\alpha' s+l-3}{\alpha' N-1}N_s^{l-1} (t) 
+q \delta_{s,1} \delta_{l,1}, 
\end{eqnarray}
with the initial condition $N_s^l(t=1)=\delta_{s,1}\delta_{l,1}$, where $\alpha'=\alpha+2$ and $q=1-p$.
Assuming the asymptotic form as $N_s^l (t)=t n_s^l=N n_s^l$ ($t \gg 1$), we obtain
\begin{eqnarray}
(\alpha' s +l+\alpha'-2)n_s^l &=& 
p (\alpha' s +l-\alpha'-2)n_{s-1}^l \\
&& +q (\alpha' s +l-3)n_{s}^{l-1}
+q \alpha' \delta_{s,1} \delta_{l,1} \nonumber.
\end{eqnarray}
Summing over $l$ gives the cluster size distribution as $n_s = \sum_l n_s^l$; 
\begin{eqnarray}
(\alpha' p s +\alpha'-2p)n_s 
&=& 
p (\alpha' s -\alpha'-2)n_{s-1} \\
&& + p \sum_l l n_{s-1}^l -p \sum_l l n_{s}^l +q \alpha' \delta_{s,1}, \nonumber 
\end{eqnarray}
which is evaluated if $\alpha \to \infty$ as 
\begin{equation} 
n_s =\frac{\Gamma(s)}{\Gamma [ s+(1+p)/p ]}n_1 \propto s^{-\frac{1+p}{p}}, 
\end{equation}
as previously reported in \cite{Kiss03,Lancaster}. 
To obtain an estimate of $n_s$ for finite offset $\alpha$, 
we assume $l^*_s=\sum_l l n_s^l/n_s$ for each $s$
to satisfy $s-1:l^*_s=p:q$.
This approximation is the same as of \cite{Pietsch}.
Then $n_s$ is obtained as 
\begin{equation}
n_s
\propto 
\frac{\Gamma[s-(\alpha'p +2)/(\alpha' p+q)+1]}{\Gamma[s+(\alpha' -p-1)/(\alpha' p+q)+1]}.
\end{equation}
Thus its asymptotic form for $s\gg 1$ is given by (\ref{nspower}) with
\begin{equation}
\tau =\frac{3+p+\alpha p+\alpha}{1+p+\alpha p} 
=\frac{\gamma+(\gamma-2) p}{1+(\gamma-2) p} \label{tau}.
\end{equation}
Note that the scaling relation (\ref{relation}) is satisfied between 
$\psi$ in Eq.(\ref{psi}) and $\tau$ in Eq.(\ref{tau}).
This result means that 
$n_s$ on the GR tree always follows power-law type, 
irrespective of $\gamma$ or $p$ ($0<p<1$). 

We may say the same in terms of the correlation {\it volume} and the correlation {\it length}.
The connectedness function $C_i(l,p)$ of node $i$ is defined as 
the probability that a randomly-chosen node, whose distance from node $i$ is $l$,
belongs to the same cluster with node $i$. 
Assuming that $C_i(l,p)$ for $p<p_{c2}$ decays as a single exponential function 
\begin{equation}
C_i(l,p) \propto e^{-l/\xi_i^l},
\end{equation}
where $\xi_i^l$ is the correlation length.
The correlation volume $\xi_i^V$ of node $i$ is given by summing the connected function as 
\begin{equation}
\xi_i^V =\sum_l n_i(l) C_i(l,p) ,
\end{equation}
where $n_i(l)$ is the number of nodes whose distance from node $i$ is $l$.
$\xi_i^V$ is the size of cluster to which node $i$ belongs.
Since $n_i(l)$ grows exponentially with $l$ for graphs with the small-world property,
it is possible that some correlation volumes diverge even when correlation length decays exponentially.
As pointed in \cite{NAG}, the critical points $p_{c1}$ and $p_{c2}$ of NAGs correspond to the points
above which the correlation volume and length diverge, respectively. 
Similarly, these grow with $p$ in a different way from each other on the GR trees: 
Eq.(\ref{psi}) indicates that the correlation volume of the initial node, $\xi_0^V$, diverges if $p>0$ ($p_{c1}=0$), 
while the correlation length $\xi_i^l$ between any node pairs 
behaves like that of one-dimensional percolations, 
$\xi_i^l \sim -1/\ln p$, and never diverges at any $p <1$ ($p_{c2}=1$). 

Note that in the MPT there is a certain open probability $p_s$ between $p_{c1}$ and $p_{c2}$,
above which the mean cluster size $\bar{s} \propto \sum_s s^2 n_s$ diverges \cite{NHpri}.
Clearly, Eq.(\ref{tau}) with $\tau=3$ gives the value $p_s=(\gamma-3)/2(\gamma-2)$, 
so that $p_s$ depends on the degree exponent $\gamma$ in contrast to $p_{c1}$ and $p_{c2}$.
For $\gamma \le 3$, $p_s$ reaches $0$ in the limit $N \to \infty$, 
so the mean cluster size diverges irrespective of $p (>0)$.
Interestingly, another expression obtained by rewriting it in terms of the moments of $k$,
$p_s=(\gamma-3)/2(\gamma-2)=\langle k \rangle/\langle k^2 -k \rangle$,
is also obtained for the percolation threshold on uncorrelated SF networks.
A similar expression for the Ising case on the GR tree has already been obtained in \cite{HN4}.

\section{numerical results}


\begin{figure}
\begin{center}
\includegraphics[width=8cm]{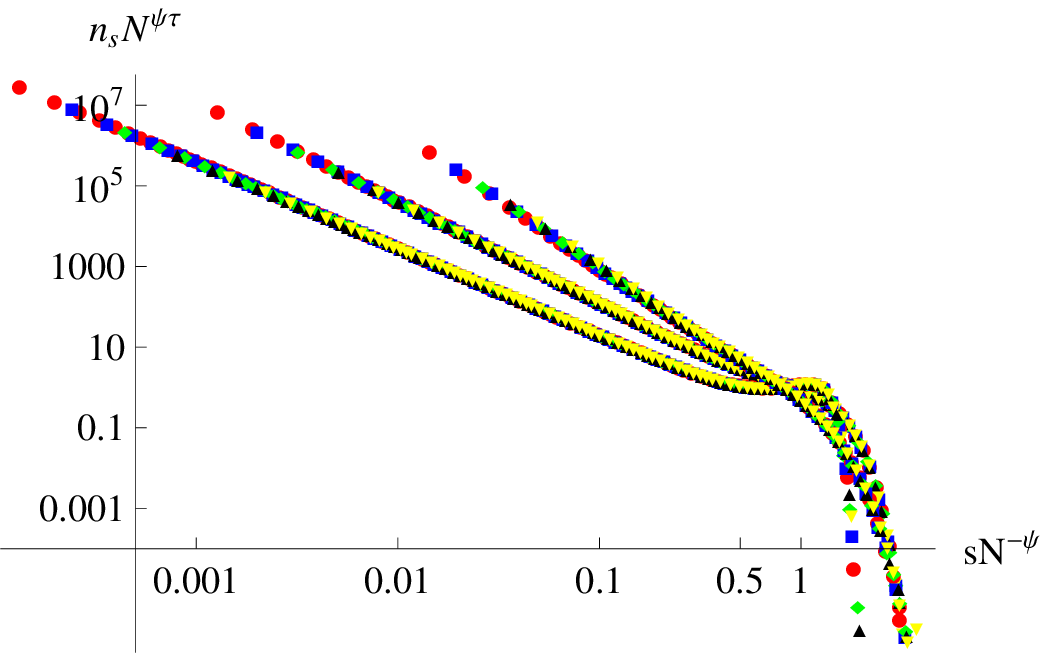}
\includegraphics[width=8cm]{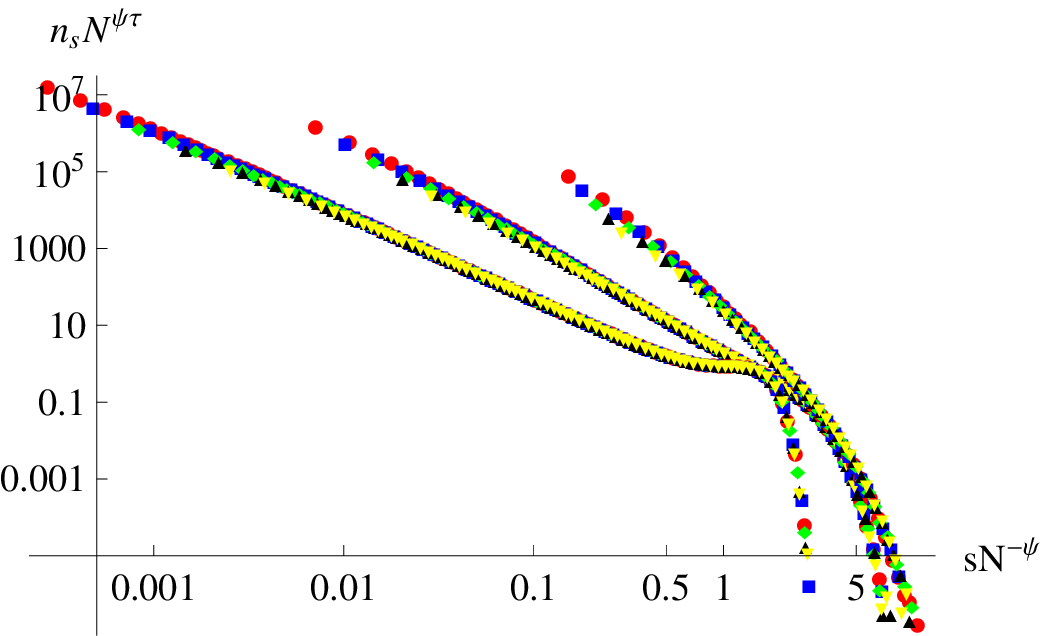}
\end{center}
\caption{
(Color online) Finite size scaling for the cluster size distribution $n_s$ 
of the GR tree with $\alpha =1$ (top panel), 100 (bottom panel). 
Each curve takes open bond probability $p=0.2, 0.5, 0.8$, from right to left.
The number of nodes is taken 
$65536$(red circle), $32768$(blue square), $16384$(green diamond), $8192$(black triangle), $4096$(yellow inverted triangle).
We obtain $\psi$ by numerical simulations, 
and determine $\tau$ by scaling relation (\ref{relation}) with numerically obtained $\psi$.
}
\label{FSSa100}
\end{figure}


In this section, we check the above analytical results by numerical calculations.
We perform the Monte-Carlo simulations for the bond percolation on the GR trees.
The number of nodes is taken from about 4096 to 65536. 
To measure the mean size of the root clusters, 
we generate 1000 graph samples from the GR tree 
and simulate the bond percolation 5000 times on each sample.
Figure \ref{psiplot} plots the fractal exponent of the root cluster $\psi$ and the analytical prediction (\ref{psi}). 
Here we evaluate $\psi$ of the system with $N$ nodes by the difference $(\ln s_0(2N)-\ln s_0(N/2))/(\ln 2N-\ln (N/2))$.
Numerically-obtained $\psi$ does not depend on $N$ except for small values of $p$ ($p \lesssim 0.2$), 
where $\psi$ tends to approach the analytical line (\ref{psi}) with increasing $N$ (not shown). 
We also measure another critical exponent $\psi_{\rm L}$ that characterizes 
the mean size of  largest clusters $s_{\rm max}(N)\propto N^{\psi_{\rm L}}$. 
The estimated value in Fig.\ref{psiplot} shows a good agreement with $\psi$, 
which indicates that the root clusters can be regarded as the largest ones, $s_0(N)\sim s_{\rm max}(N)$.
To check $n_s$, we assume a finite size scaling form \cite{NAG};
\begin{eqnarray}
n_s(N) = N^{-\psi \tau} f( s  N^{-\psi} ), 
\end{eqnarray}
where the scaling function behaves as
\begin{equation}
f(x)\sim\left\{
\begin{array}{cll}
\text{rapidly decaying func.} & \text{for} & x\gg1,\\
x^{-\tau} & \text{for} & x\ll 1.
\end{array}
\right.
\end{equation}
Here we also assume the scaling relation (\ref{relation}).
Figure \ref{FSSa0} shows our finite size scaling for $n_s$ on the GR tree with $\gamma=3$.
Our scaling with several sizes are quite well fitted in a wide range of $p$.
We also find similar results for various values of $\gamma$ 
as shown in Fig.\ref{FSSa100}.
We consider the scaling relation (\ref{relation}) is quite general. 
If $n_s\propto s^{-\tau}$ holds asymptotically,
{\it a natural cutoff} $s_{\rm max}(N)$ of the cluster size distribution 
(a natural cutoff of the degree distribution was introduced in \cite{DorogoCutoff}) is given as 
\begin{equation}
N \int_{s_{\rm max}(N)}^{\infty} n_s ds \simeq 1 \to s_{\rm max}(N) \propto N^{\frac{1}{\tau-1}}. \label{naturalcutoff}
\end{equation}
Then Eq.(\ref{relation}) follows with replacing $\psi_{\rm L}$ by $\psi$. 

Finally we note that a standard finite size scaling analysis does not work to
determine $p_{c1}$ and $p_{c2}$ by using the data in the critical phase (see \cite{NAG2} for the similar argument of the enhanced binary tree).
Suppose that in the critical phase $s_{\rm max} \propto N^{\psi_{\rm L}}$ 
holds with an increasing analytic function of $\psi_{\rm L}(p)$
for $p_{c1}<p<p_{c2}$.
Then one can expand $\psi_{\rm L}(p)$ around any $p^*$ in the critical phase so that
$\psi_{\rm L}(p)-\psi_{\rm L}(p^*) \propto p-p^*$, which leads us to
\begin{equation}
s_{\rm max} N^{-\psi_{\rm L}(p^*)}=g(\log N(p-p^*)),
\end{equation}
where $g(.)$ is a "scaling function" around $p^*$.
This means that the scaled parameter $p^*$ is arbitrary as long as it is in the critical phase
and its boundary $p_{c1}$ and $p_{c2}$ is hardly determined from this type of analysis
for practical use.

\section{summary}

Our results indicate that the bond percolation on the GR tree has a critical phase for $0<p<1$, 
and the critical behavior is similar to that observed in a NAG \cite{NAG}. 
The critical phase is characterized by a power-law behavior of $n_s$ with varying exponent.
The same property is observed 
on the decorated (2,2)-flower \cite{Berker09} and the Hanoi networks \cite{Boettcher}.
After the submission of this paper, Sato and the authors calculated 
$\psi$ and $\tau$ of the decorated (2,2)-flower by a generating functional approach to confirm 
that the phase is indeed critical everywhere \cite{SHN}.
As already mentioned, 
the critical phase is caused by opening a gap between the points 
at which the correlation volume and correlation length start to diverge.
What is the structural factor of networks for such a separation to occur?
Finding the answer is, we believe, an essential step in understanding usual and unusual phase transitions on complex networks.

We thank T. Nogawa for helpful discussions.
This work was supported by the 21st Century Center of Excellence (COE) program entitled "Topological Science and Technology", Hokkaido University.

\end{document}